\documentclass[nofootinbib,superscriptaddress,twocolumn,aps,prc,showpacs,10pt]{revtex4-1}
\usepackage{verbatim}
\usepackage{amsmath}
\usepackage{amssymb}
\usepackage{graphicx}
\usepackage{mathrsfs}
\usepackage{braket}
\usepackage{color}

\begin{document}

\title{Combining phase-space and time-dependent reduced density matrix approach to describe the dynamics of interacting fermions}
\author{Thomas Czuba} \email{czuba@ipno.in2p3.fr}
\affiliation{Institut de Physique Nucl\'eaire, IN2P3-CNRS, Universit\'e Paris-Sud, Universit\'e Paris-Saclay, F-91406 Orsay Cedex, France}
\author{Denis Lacroix} \email{lacroix@ipno.in2p3.fr}
\affiliation{Institut de Physique Nucl\'eaire, IN2P3-CNRS, Universit\'e Paris-Sud, Universit\'e Paris-Saclay, F-91406 Orsay Cedex, France}
\author{David Regnier} 
\affiliation{Institut de Physique Nucl\'eaire, IN2P3-CNRS, Universit\'e Paris-Sud, Universit\'e Paris-Saclay, F-91406 Orsay Cedex, France}
\affiliation{Centre de math\'ematiques et de leurs applications, CNRS, ENS Paris-Saclay, Universit\'e Paris-Saclay, 94235, Cachan cedex, France.}
\author{Ibrahim Ulgen}
\affiliation{Physics Department, Faculty of Sciences, Ankara University, 06100 Ankara, Turkey}
\author{Bulent Yilmaz}
\affiliation{Physics Department, Faculty of Sciences, Ankara University, 06100 Ankara, Turkey}

\date{\today}

\begin{abstract}
The possibility to apply phase-space methods to many-body interacting systems might provide accurate descriptions of correlations 
with a reduced numerical cost. For instance, the so--called stochastic mean-field phase-space approach, where the complex dynamics of interacting fermions is replaced 
by a statistical average of mean-field like trajectories  is able to  grasp some correlations beyond the mean-field. 
We explore the possibility to use alternative equations of motion in the phase-space approach. Guided by the BBGKY hierarchy, 
equations of motion that already incorporate part of the correlations beyond mean-field are employed along each trajectory. 
The method is called Hybrid Phase-Space (HPS)
because it mixes phase-space techniques and the time-dependent reduced density matrix approach. The novel approach is applied to the one-dimensional 
Fermi-Hubbard model. We show that the predictive power is improved compared to the original stochastic mean-field method. In particular, in the weak-coupling regime, the results of the HPS theory can hardly be distinguished from the exact solution even for long time.        
\end{abstract}


\maketitle

\section{Introduction}

The accurate description of the evolution of interacting fermions is an extremely challenging problem when the number of particles 
increases. One of the difficulties is the number of degrees of freedom (DOFs) to be followed in time that scales exponentially with the number 
of particles. A natural way to reduce the complexity is to assume that some DOFs are more relevant than others and to follow in time only these DOFs. 
A typical illustration of this strategy is the Time-Dependent Hartree-Fock (TDHF) approach where one-body DOFs are assumed to contain the relevant information on the 
system evolution. This reduction of information is evident when we consider as a starting point the  Bogolyubov- Born-Green-Kirkwood-Yvon (BBGKY) hierarchy \cite{Bog46, Bor46, Kir46,Cas90,Gon90,Sch90}. Then, the TDHF theory is recovered by assuming that two-body, three-body, $\dots$ DOFs can all be written in terms of the one-body density (see for instance \cite{Bon16}). The BBGKY approach also provides strong guidance to go beyond the mean-field approximation by including gradually higher order effects related to two-body,
three-body, $\dots$ DOFs. This has led to a variety of approaches that can be referred to as the Time-Dependent Reduced Density Matrix (TDRDM). More precisely such approach 
can be called TD$n$RDM where the $n$ is the maximal order of the reduced density matrix that is considered in the description. Solving the TD$n$RDM with $n>1$, even today, remains a complicated numerical task and the approximation used to truncate the BBGKY hierarchy has to be analyzed with special care (see for instance the recent discussions 
in \cite{Lac15b,Lac17} and references therein).
        
Phase-space approaches offer an alternative scheme allowing to describe correlations beyond mean-field.  
In these approaches, a complex dynamical problem is replaced by a set of simpler dynamical evolutions. Then, the complexity of the dynamics can eventually be 
described by a proper weighted average over the simpler evolutions \cite{Gar00}. 
An example of such approach that has been applied in bosonic interacting systems with some success is the Truncated-Wigner 
Approximation (TWA) \cite{Sin02}. Less attempts have been made to develop and apply Phase-Space approaches in Fermi systems. We mention the so-called 
Stochastic Mean-Field (SMF) theory that was proposed already some times ago \cite{Ayi08} and tested also with some success 
\cite{Lac12,Lac13,Lac14b} (for a review see \cite{Lac14a}).  Another approach, that turns out to be rather close to the SMF technique, is the fermion-TWA (f-TWA) of 
Ref. \cite{Dav17}.   

In the SMF phase-space approach proposed in Ref. \cite{Ayi08}, the initial quantum fluctuations in many-body space are mimicked by a Gaussian statistical ensemble
of initial one-body densities. Then, each initial condition follows a TDHF like trajectory
that plays the role of the "simple" evolution.  We already have shown in Refs. \cite{Yil14,Ulg19} that the approach can benefit from relaxing the Gaussian approximation 
for the initial statistical ensemble. Our aim here is to explore if alternative equations of motion for individual trajectory can be proposed that would improve the predictive power 
of this phase-space method.  
To further progress, we realized that a more careful analysis of the connection between the phase-space 
approach proposed in Ref. \cite{Ayi08} and the BBGKY hierarchy should be made. For this reason, we start the discussion below by recalling 
basic aspects of this hierarchy that will be useful later. Then, we propose a novel phase-space 
approach inspired from both SMF and BBGKY that we called Hybrid Phase-Space (HPS). We show that it indeed improves the 
description of interacting systems.    

\section{Many-body dynamics: BBGKY versus Phase-space methods}

\subsection{BBGKY and truncation schemes}

In the present article, we consider a general two-body Hamiltonian written in the second quantized form as:
\begin{eqnarray}
H = \sum_{ij} t_{ij} \hat{a}^\dagger_i \hat{a}_j + \frac{1}{4} \sum_{ijkl} \widetilde v_{ijkl }  \hat{a}^\dagger_{i} \hat{a}^\dagger_{j}
\hat{a}_{l} \hat{a}_{k}.
\end{eqnarray} 
Here $\widetilde v_{12}$ denotes the antisymmetric matrix elements \footnote{Through this paper we will use 
the notations \cite{Lac14a} where the indices refer to the particle to which the operator applies. For instance 
$\langle ij | \widetilde v_{12} | kl \rangle = \langle ij |  v_{12} \left(1 - P_{12} \right)| kl \rangle =  V_{ij,kl} -  V_{ij,lk}$, where $P_{12}$ is such that $P_{12}| kl\rangle = | lk\rangle$.}.
The initial condition is given in terms of the N-body density matrix 
$D(t_0)$ that contains the information on the initial state of a set of independent or correlated 
fermions. Our aim is to provide an accurate description of the system evolution for time $t>t_0$. 
The exact solution to this problem can be obtained by solving the Liouville-von Neumann equation given by:
\begin{eqnarray}
i\hbar \dot D(t) &=& \left[ H ,  D(t) \right], \label{eq:Liouv}
\end{eqnarray}
where $\dot D(t)$ denotes the time-derivative of $D(t)$. In many realistic situations, the direct use of Eq. (\ref{eq:Liouv}) 
is intractable due to the number of components of $D(t)$, that are directly connected to the number of DOFs  to follow in time.  A standard way to reduce the complexity is to assume 
that there is a hierarchy in the importance of selected degrees of freedoms compared to others. Often, 
the one-body DOFs are assumed to be more important than two-body DOFs that are both supposed to be more important than three-body DOFs and so on 
and so forth.   The usual method to focus on the k-body DOFs consists in introducing the k-body reduced density matrix (kRDM), defined through:
\begin{eqnarray}
\langle k'\cdots 1'| R_{1\cdots k} | 1 \cdots k \rangle = \langle \hat a^\dagger_1 \cdots \hat a^\dagger_k \hat a_{1'} 
\cdots \hat a_{k'}   \rangle. \nonumber
\end{eqnarray}  
In the following, we will mainly focus on the one-, two- and three-body density matrices, denoted respectively by $R_1$, $D_{12}$ and $T_{123}$. 
Assuming that the number of particles in the system is $N$, these densities are linked to each other through the partial trace relations:
\begin{eqnarray}
(N-1) R_1 &=& {\rm Tr}_{2} D_{12}, ~~(N-2)  D_{12} =  {\rm Tr}_{3} T_{123}. \label{eq:partialtrace}
\end{eqnarray}

Starting from Eq. (\ref{eq:Liouv}), 
one can derive the well-known BBGKY hierarchy of equations of motion (EOMs) \cite{Bog46, Bor46, Kir46,Cas90,Gon90,Sch90}, showing that
the kRDM evolution is coupled to the (k+1)RDM. For the present discussion, we will only need the two first equations of the hierarchy that are given 
respectively by: 
  \begin{eqnarray}
i \hbar \dot R_1 &=& \left[ t_1 , R_1 \right] + \frac{1}{2} { \rm Tr}_2 
\left[  \tilde v_{12}, D_{12} \right] , \label{eq:R1} 
\end{eqnarray}    
and 
\begin{eqnarray}
i \hbar \dot D_{12} &=& [H_{12}, D_{12}] 
+ \frac{1}{2} {\rm Tr}_3 \left[ (\tilde v_{13} + \tilde v_{23}) , T_{123} \right], \label{eq:R12}
\end{eqnarray}
with $\displaystyle H_{12} = t_1 + t_2 + \frac{1}{2} \tilde v_{12}$. The BBGKY hierarchy has been and is still   
a continuous source of inspiration to obtain approximate treatments of the N-body dynamical problem. The standard  
strategy is to truncate the hierarchy at a given order $k$ while using a prescription for the densities of orders 
higher than $k$ so that they can be written as a functional of lower orders reduced densities. The simplest example is
the Time-Dependent Hartree-Fock (TDHF) theory that is recovered from  Eq. (\ref{eq:R1}) assuming that the 
2RDM is given by $D_{12} = R_1 R_2 (1-P_{12})$. 
The resulting equation then writes:
 \begin{eqnarray}
i \hbar \dot R_1 &=& \left[ t_1 , R_1 \right] + { \rm Tr}_2 
\left[  \tilde v_{12}, R_{1} R_{2} \right]  \label{eq:TDHF1}  \\
&\equiv& \left[ h_1[R], R_1 \right],  \label{eq:TDHF2}
\end{eqnarray}     
where $h_1[R] = t_1 + {\rm Tr}_2 [ \tilde v_{12}  R_{2}]$ denotes the mean-field. Staying at the mean-field level is
generally not sufficient to describe interacting systems
and most often two-body or higher correlations 
between particles should be included explicitly. For instance, large efforts are devoted to obtain closed EOMs 
between the 1RDM and 2RDM or solely for the 2RDM matrix \cite{Yas97,Maz00,Toh10,Toh14a}. One delicate issue is the prescription used to truncate the BBGKY 
hierarchy that might strongly impact the quality of the results \cite{Akb12,Toh19}. Related to this issue is the possible breakdown of some important conservation 
laws when writing the 3RDM in terms of the 2RDM and 1RDM \cite{Sch90}. We note that an interesting solution to this problem was recently given with 
the purification technique proposed in Refs. \cite{Lac15b,Lac17}.

\subsection{Phase-Space approach applied to Fermi systems}
\label{sec:phasespace}

 In the present work, we will call "Phase-Space" approach a technique where a complex quantum dynamical problem is replaced by an ensemble of simpler dynamical 
problems with a statistical ensemble of initial conditions.  The statistical properties of the initial ensemble are chosen at best to reproduce the initial properties of the complex 
system to be simulated. As mentioned in the introduction, very few practical phase-space theories to simulate fermionic interacting systems have been proposed so far \cite{Ayi08,Dav17}. 

Here, we will use the SMF theory that we are familiar with as a starting point. In this approach, a statistical ensemble of one-body densities is considered. Each realization 
of the initial statistical ensemble, denoted by $R^{(n)}_1$, where $(n)$ labels  the event,  is then evolved assuming that the 1RDM follows a mean-field like trajectory that is independent from the other trajectories 
\begin{eqnarray}
i\hbar \dot R^{(n)}_1 &=& \left[ h_1[R^{(n)}], R^{(n)}_1 \right]. \label{eq:eomsto}
\end{eqnarray}  

There are two important ingredients in this phase-space method: 
\begin{itemize}
  \item[(a)] the statistical properties of the initial 
ensemble,

  \item [(b)] the choice of the equation of motion for the 1RDM.
  \end{itemize}
In \cite{Ayi08,Dav17}, Gaussian probabilities are assumed for the matrix elements of the 1RDM 
such that their  first and second moments match the one of the initial complex state one wants to describe. Let us for instance assume that the initial state is a
simple independent particle state at zero or finite temperature. Then, the information on the system is contained in its one-body density matrix that is given
in the natural basis denoted by ${\phi_\alpha}$ by $R_1 (t_0)= \sum_\alpha | \phi_\alpha (t_0)\rangle n_\alpha(t_0) \langle \phi_\alpha (t_0) |$. To reproduce the properties of the initial state,  
it was shown in Ref. \cite{Ayi08} that the initial ensemble of 1RDM should fulfill the following conditions at initial time (omitting $t_0$ for compactness):

\begin{eqnarray}
\left\{
\begin{array}{l}
\overline{R^{(n)}_{\alpha \beta}} = \delta_{\alpha \beta} n_\alpha,  \\
\\
\overline{ \delta  R^{(n)}_{\alpha \beta}\delta  R^{(n)}_{\gamma \delta} } =
\frac{1}{2} 
\delta_{\alpha \delta}\delta_{\beta \gamma}\left[ n_\alpha (1-n_\beta) + n_\beta (1 - n_\alpha)\right],
\end{array}
\right.  \label{eq:SMFfluc}
\end{eqnarray}  
where $ \delta  R^{(n)}_{i j} = R^{(n)}_{i j}  - \overline{R^{(n)}_{ij}}$ and where $\overline{X^{(n)}}$ denotes here the statistical average. 
An important aspect of the SMF theory is that the original quantum framework is replaced by a statistical  treatment. For instance, 
any one-body observable $O$ becomes a fluctuating quantity ${O^{(n)}}$ given at time $t$ by:
\begin{eqnarray}
O^{(n)}(t) & = & {\rm Tr}(O R^{(n)}(t)) = \sum_{ij} O_{ij}  R^{(n)}_{ji}(t). \nonumber 
\end{eqnarray}  
The fluctuations properties are then obtained using classical statistical average over the events, for instance the mean value is given by 
$\overline{O^{(n)}(t)} = {\rm Tr}(O \overline{R^{(n)}(t)})$ while the 
second central moment, denoted by $\Sigma^2_O(t)$, is obtained through:
\begin{eqnarray}
\Sigma^2_O(t) &=& \overline{O^{(n)}(t) O^{(n)}(t)} - \overline{O^{(n)}(t)}^2   \nonumber \\
&=& \sum_{ij, kl} O_{ij} O_{kl} \overline{ \delta R^{(n)}_{ji}(t) \delta R^{(n)}_{lk}(t) }.
\end{eqnarray}
An important property resulting from Eqs. (\ref{eq:SMFfluc}) is that the statistical average of the mean value and fluctuations matches the quantum mean and 
fluctuations of the quantum problem at initial time.  

Applications of the SMF approach have shown several appealing features. One of the attractive aspects is that its predictive power 
can compete for instance with the TD2RDM approach while only requiring the propagation of the one-body density.  In general, it was found 
that the approach is highly competitive when the interaction between particles is not too strong and, whatever the strength of the 
interaction, it properly describes the short time evolution as well as the average asymptotic behavior. An illustration of application 
is given below. 

\subsubsection{Illustration of application in the 1D Fermi-Hubbard model}

In order to illustrate the SMF predictive power, we follow Ref. \cite{Lac14b} and apply the approach to the 1D Fermi-Hubbard model. 
The reason why we specifically focused on this model is because it was one of the most difficult to describe within the phase-space approach 
compared to other applications \cite{Lac12,Lac13,Yil14} and, even in the weak-coupling, the long-time evolution was impossible to reproduce. Therefore, it is a perfect test-bench
for quantifying the departure from the exact evolution and/or for testing possible improvements beyond SMF.     

In this model, the Hamiltonian describes interacting fermions of spin $\sigma$ that can move in a set of doubly-degenerated sites labelled by $i$ and associated
to creation/annihilation operators $(\hat{c}^\dagger_{i\sigma}, \hat{c}_{i\sigma})$. The Hamiltonian is given here by 
\begin{eqnarray}
H & = & - J \sum_{i,\sigma} \left\{ \hat{c}^\dagger_{i\sigma}  \hat{c}_{i+1 \sigma} (1-\delta_{iN_s}) + \hat{c}^\dagger_{i\sigma}  \hat{c}_{i-1 \sigma} (1-\delta_{i1}) \right\} \nonumber \\
&&+ U \sum_i \hat{c}^\dagger_{i,\uparrow} \hat{c}^\dagger_{i \downarrow} \hat{c}_{i \downarrow} \hat{c}_{i \uparrow}, \label{eq:Hubbard_Hamiltonian}
\end{eqnarray} 
where we use sharp boundary conditions. The model can be interpreted as a schematic Hamiltonian describing interacting particles on a lattice 
where particles can tunnel from one site to neighboring ones, the tunneling being described in an effective way by the $J$ term. The $U$ term acts 
as a local Coulomb interaction between 2 electrons that are on the same site. For more detailed interpretation of the Hubbard 
model, see for instance \cite{Jak98,Gre02}.  The exact solutions, that are shown below, are obtained here by directly solving the coupled 
equations between the coefficients of the decomposition of the time-dependent state on a full many-body basis using the spin symmetry of the initial state 
(see discussion in appendix \ref{sec:compdetHM}).     

Following Ref. \cite{Lac14b}, we consider the case where the number of particles $N$ is equal to the number of sites $N_s$ (assumed to be even in the following) 
and suppose that all particles are initially located on
one side of the mesh.  The initial state then corresponds to a Slater determinant with occupation numbers, denoted by $n_{i\sigma}=1$ if $i<N_s/2$ and $0$ otherwise. 
These occupation probabilities are related to the one-body density through  $n_{i\sigma} \equiv R^{\sigma \sigma}_{ii}$, where we used the notation $R^{\sigma \sigma'}_{ij} = \langle \hat c^\dagger_{j\sigma'} \hat c_{i\sigma}\rangle$. 
The mean-field equation of the one-body density components are given in appendix  B of 
\cite{Lac14b} as well as the statistical properties of the initial ensemble of one-body density matrices used when applying the SMF approach.
For the sake of completeness, the SMF equation for the Fermi-Hubbard model are recalled in appendix \ref{sec:compdetHM}. 

\begin{figure}[!h]
    \includegraphics[width=1.0 \linewidth]{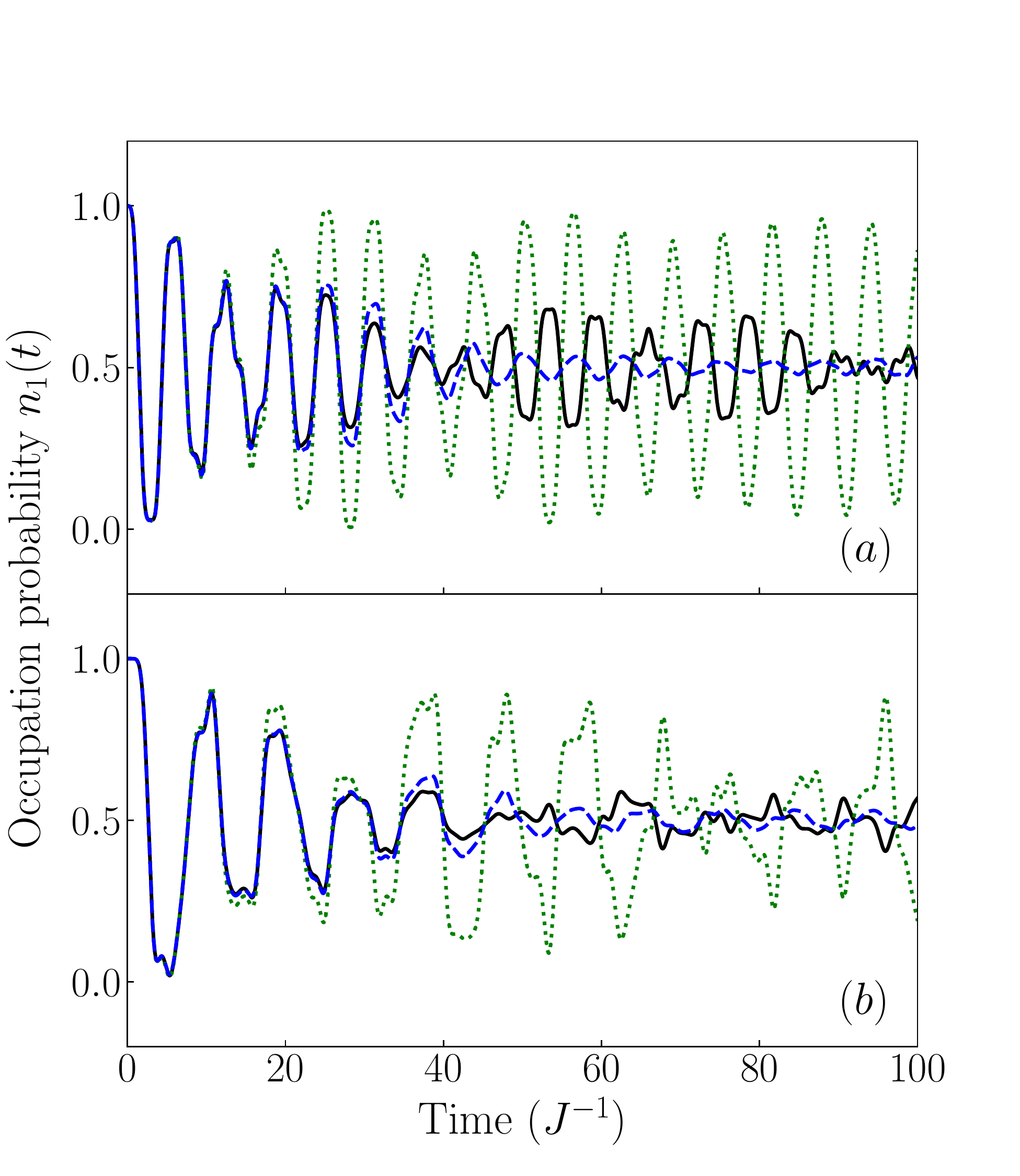}
    \caption{Time evolution of the occupation probability of the leftmost site denoted by $n_1(t)=R^{\sigma\sigma}_{11}(t)$ 
    for a ratio $U/J =0.1$ and (a) $N= N_s = 4$ or (b) $N= N_s = 8$. In both panels, the exact solution is displayed by a black solid line, 
    the TDHF solution is given by a green dotted line and the average over the SMF phase-space trajectories is given by a blue dashed line. The average occupation number is obtained here by averaging over 10000 trajectories. Note that here we have $n_{1\uparrow}(t)= n_{1\downarrow} (t)= n_1(t)$ and we simply omit the spin.} 
    \label{fig:48mfvssmf}
\end{figure}

For small number of sites, the problem can be solved exactly and can be confronted to approximate treatments. We compare in Fig. \ref{fig:48mfvssmf}, the 
exact solution obtained for 4 (resp. 8) particles on 4 (resp. 8) sites  with both the mean-field and stochastic mean-field solution using the Gaussian assumption 
for the initial statistical ensemble and for a coupling strength $U/J =0.1$. 
In the following, we will use the convention $\hbar=1$ and time will be given 
in $J^{-1}$ units.   

We clearly see that a significant improvement in the description of the evolution is achieved in the SMF approach compared to the 
TDHF case. For instance, the damping of $n_1(t)$ is remarkably well reproduced up to $t \simeq 40 J^{-1}$ and deviation from the exact solution 
is only observed for long time evolution. In general, it is found \cite{Lac14a} that the predictive power of SMF is rather good in the weak coupling regime 
and degrades when the coupling increases. In addition, while it uses only mean-field like EOMs, it is found to be able to compete 
with other approaches like those based on the truncation of the BBGKY hierarchy we have discussed previously. We have shown in Ref. \cite{Yil14} and more recently 
in \cite{Ulg19}, that the approach can be sometimes further improved by relaxing the Gaussian approximation on the initial fluctuations.  In the specific case of the Hubbard         
model, we have tried to replace the initial Gaussian ensemble by a  two-point distribution as proposed in \cite{Ulg19} but the improvements were marginal. Below, we 
propose a novel approach that combines the SMF with the BBGKY hierarchy truncation technique.

\subsubsection{k-body density matrix in SMF and BBGKY like hierarchy on symmetric moments}

As noted in Ref. \cite{Lac15},  an explanation of the SMF success is that this approach is equivalent  to solve an untruncated infinite set 
of coupled equations of motion on the moments defined as:
\begin{eqnarray}
M_{1 \cdots k} & = &  \overline{R^{(n)}_{1} \cdots R^{(n)}_{k}} \equiv \overline{M^{(n)}_{1 \cdots k}}.  \label{eq:momdens} 
\end{eqnarray}
As explained in the appendix \ref{sec:appfromsmfbbgky}, these moments play a special role in the SMF approach in many respects that we recall below:
\begin{itemize}
  \item First of all, $M^{(n)}_{1 \cdots k}$  does contain the information on k-body correlations between observables. Indeed, let us consider a set of one-body observables 
  $\{ O^\alpha \} $ with $\alpha=1, \cdots, k$. In the phase-space method, we have:
 \begin{eqnarray}
\overline{O^{1} \cdots O^{k}} &=& \sum_{ij, \cdots , mn} O^{1}_{ij} \cdots O^{k}_{mn}   \overline{R^{(n)}_{ji} \cdots R^{(n)}_{nm}} \nonumber \\
&=& \sum_{ij, \cdots , mn} O^{1}_{ij} \cdots O^{k}_{mn} M_{j \cdots n, i \cdots m}.  \nonumber
\end{eqnarray}    
  \item Similarly to the set of density matrices defined by Eq. (\ref{eq:partialtrace}), the moments are linked with each other through a partial trace relation that holds event-by-event:
\begin{eqnarray}
{\rm Tr}_{k+1} M^{(n)}_{1 \cdots k+1} (t) &=& N \times  M^{(n)}_{1 \cdots k}(t), \label{eq:tracemom}
\end{eqnarray}    
where we used the fact that $ {\rm Tr}R^{(n)}_1(t)= N$ for all trajectories and at all time. Since this property holds for each event, it is also valid in average.  
  \item The SMF phase-space approach can also be interpreted as the following mapping at initial time:
  \begin{eqnarray}
\langle \{ \hat N_{ij} \}_+ \rangle \longrightarrow  \overline{R^{(n)}_{ij}}, \nonumber \\
\langle \{ \hat N_{ij}, \hat N_{kl} \}_+ \rangle \longrightarrow   \overline{R^{(n)}_{ij} R^{(n)}_{kl}} , \nonumber \\
\end{eqnarray} 
where $\hat N_{ij}= \hat a^\dagger_j \hat a_i$ and where $\langle \{ . , \cdots, . \}_+ \rangle$   denotes the quantum expectation value of the fully symmetric moments (for further details 
see appendix \ref{sec:appfromsmfbbgky}).  In the quantum problem, these quantum symmetric moments contain the same information as the density matrices. This is illustrated 
for the one-, two- and three-body densities with Eqs. (\ref{eq:d1sym}-\ref{eq:d3sym}).  For a Gaussian distribution of the initial fluctuations, the mapping is exact at initial time only for the 
first two moments and only approximate for higher moments. From this mapping, one can also define in a clean way the equivalent to the density matrices within the SMF framework. The expression of the event-by-event two-body and three-body density matrices are respectively given by Eq. (\ref{eq:dsmf}) and (\ref{eq:tsmf}). In particular, consistently with the Gaussian approximation, we again deduce that the average one- and two-body densities matches the exact quantum densities at initial time.    
\item Finally, starting from the TDHF equation of motion on $R^{(n)}_1$ and using the explicit form of the mean-field Hamiltonian, it is rather simple to show \cite{Lac15} that, event by event, the set of moments follow a set of coupled equations where at a given order $k$, the moment $M^{(n)}_{1 \cdots k}(t)$  is coupled to the moment  $M^{(n)}_{1 \cdots k+1}(t)$. 
Then, by averaging over the events, an equivalent hierarchy is obtained on the average moments. For the following discussion, we give the explicit form of the first two equations of the hierarchy. The first equation reads:
  \begin{eqnarray}
i \hbar \dot R^{(n)}_1 
&=&  \left[ t_1 , R^{(n)}_1 \right] +  { \rm Tr}_2  \left[  \tilde v_{12}, M^{(n)}_{12} \right]   \label{eq:mom1},
\end{eqnarray}  
while the equation on the second moment is given by:
\begin{eqnarray}
i\hbar \frac{d}{dt} M^{(n)}_{12}(t) &=& \left[ t_{1} + t_{2} , M^{(n)}_{12} \right] \nonumber \\
&+&  {\rm Tr}_3 \left[ (\tilde v_{13} + \tilde v_{23}) ,  M^{(n)}_{123}  \right]. \label{eq:mom2}
\end{eqnarray}       
These equations and their average counterparts illustrate how non-trivial effects beyond the mean-field picture are incorporated within SMF. 
Taking the average over trajectories, we readily obtain the first two equations of the hierarchy coupling $R_1$ to $M_{12}$, $M_{12}$ to $M_{123}$, and so on and so forth.
\end{itemize}  

\section{Hybrid phase-space method}
\label{sec:hybrid}

The clear advantage of the SMF theory highlighted above is its predictive power despite the fact that only the mean-field machinery
is involved. We indeed 
recurrently observed that the approach can compete with other techniques where two-body DOFs are explicitly 
evolved in time. The approach is however not exact and leads to deviations with the exact results, for instance for long time evolution even in the weak coupling regime (see Fig. \ref{fig:48mfvssmf}). Its predictive power degrades when the strength of the two-body interaction increases. 

The building blocks of the approach are the two assumptions made for the items (a) and (b) discussed in      
section \ref{sec:phasespace}, respectively the Gaussian assumption for the initial statistical ensemble and the mean-field like dynamics
of $R_1^{(n)}$ along each path.    
In recent years, we have already explored the possibility to relax the Gaussian approximation for the initial probabilities in Refs. \cite{Yil14,Ulg19}. Our 
conclusion is that, although a systematic way of deciding the form of the initial  probabilities is still missing, non-Gaussian probabilities that are better optimized 
to reproduce the initial system can lead to non-negligible improvements in the description of its evolution. 
Unfortunately, the alternative prescription proposed in Ref. \cite{Ulg19} leads to only small improvement compared to the Gaussian case for the Fermi-Hubbard model.
 
The original motivation of the present work was to use the BBGKY hierarchy as a guidance to propose an equation of motion for $R^{(1)}$ that could provide 
an alternative to the mean-field like equation used in SMF and eventually 
increase the predictive power. 
A first hint in this direction was given in Ref. \cite{Puc15,Ori17} for bosonic like systems where higher order equations of 
the BBGKY hierarchy were used to extend the TWA approach and leads to an improved description of the evolution. It turns 
out that the method we propose below not only reaches the goal for item (b) but might also be useful to better describe the initial 
state. 

\subsection{Critical analysis of the standard SMF approach and its connection with the BBGKY hierarchy}

The strategy we  follow  to change the EOMs used in SMF is to make connection between the hierarchy 
of equation on the moments obtained from the average SMF evolution and the BBGKY hierarchy obtained for the k-body densities
in the quantum many-body problem. As we have seen in the SMF approach, the hierarchy of dynamical equations on moments is relatively simple.
In parallel, in the BBGKY hierarchy, the set of equations on the densities are relatively simple too. Unfortunately, the opposite is not true. Starting from the 
SMF average moments, we can obtain the corresponding average density (see for instance Eqs. (\ref{eq:dsmf}-\ref{eq:tsmf}). The expressions and as a consequence 
the equation of motion for the average density are complex. On the other hand, starting from the BBGKY hierarchy, one can express the quantum 
symmetric moments in terms of the densities (see discussion in the appendix \ref{sec:appfromsmfbbgky}), but in this case, it is the EOMs on the quantum 
moments that become rapidly extremely complex.   
This complexity has prevented us from finding a systematic constructive way to improve the EOMs to be used in the phase-space approach.  
Below, we propose a more pragmatic approach. 

\subsection{Hybrid Phase-Space (HPS) method guided by the BBGKY hierarchy}

Besides the Gaussian assumption for the initial noise, the first evident source of errors in SMF can be seen by taking the average evolution of $R^{(n)}_1$. 
Indeed, taking the average of Eq. (\ref{eq:mom1}) and using the relation between the average moment $\overline{M^{(n)}_{12}}$  
and the average density $\overline{D^{(n)}_{12}}$ obtained by averaging Eq. (\ref{eq:dsmf}), we immediately see that the evolution does not match 
the first BBGKY equation given by (\ref{eq:R1}).     

Based on this observation and in order to improve the phase-space approach, we will force the event-by-event one-body  
evolution (EOM) to take the form
\begin{eqnarray}
i \hbar \dot R^{(n)}_1 &=&  \left[ t_1 , R^{(n)}_1 \right] +  \frac{1}{2} { \rm Tr}_2  \left[  \tilde v_{12}, {\cal D}^{(n)}_{12} \right]. \label{eq:hybrid1}
\end{eqnarray}  
Although we might be tempted to interpret ${\cal D}^{(n)}_{12}$  as a fluctuating two-body density, for the moment, the only constraint we impose is    
that it has some properties of the exact two-body density matrix (antisymmetry, hermiticity).  We also assume that  ${\cal D}^{(n)}_{12}$ evolves according to an equation of motion similar to the second BBGKY equation that is given by:
\begin{eqnarray}
i \hbar \dot {\cal D}_{12}^{(n)} &=& \left[ H_{12} , {\cal D}_{12}^{(n)} \right] +  \frac{1}{2} {\rm Tr}_3 \left[ (\tilde v_{13} + \tilde v_{23}) , {\cal T}^{(n)}_{123} \right], \label{eq:hybrif2temp} 
\end{eqnarray}
where ${\cal T}^{(n)}_{123}$ is for the moment an intermediate quantity that has the same properties as the three-body density matrix. Obviously, if at all time 
we have $\overline{{\cal D}^{(n)}_{12}}(t) = D_{12} (t)$ and $\overline{{\cal T}^{(n)}_{123}}(t) = T_{123} (t)$ where $D_{12} (t)$  and $T_{123} (t)$ are the exact quantum density, then 
the averages of the above two equations of motion match the exact evolution. However, constraining the one-, two- and three-body fluctuating quantities to match 
in average the exact evolution is an open problem by itself.  

A slightly simpler task, that follows the spirit of the SMF approach, is to impose constraints only  at initial time, and more precisely, our goal is to impose:
\begin{eqnarray}
\left\{
\begin{array}{l}
\overline{ R^{(n)}_{1}}(t_0) = R_{1} (t_0),    \\
\\
\overline{{\cal D}^{(n)}_{12}}(t_0) = D_{12} (t_0), \\
\\
\overline{{\cal T}^{(n)}_{123}}(t_0) = T_{123} (t_0).
\end{array}
\right. 
 \label{eq:initrdt}
\end{eqnarray}   
The first two constraints are already fulfilled in the original SMF formulation \cite{Ayi08} using the statistical properties 
given by Eq. (\ref{eq:SMFfluc}) and the Gaussian assumption for the initial statistical ensemble. However, with this Gaussian approximation, $\overline{{\cal T}^{(n)}_{123}}(t_0)$ obtained by averaging Eq. (\ref{eq:tsmf})   does not match $T_{123} (t_0)$, even starting from a pure Slater determinant state.

Solving  Eq. (\ref{eq:hybrif2temp}) also requires to have the equation of motion for the quantity ${\cal T}^{(n)}_{123}(t)$. To avoid this, we simply close the EOM between 
$R^{(n)}_{1}(t)$ and ${\cal D}^{(n)}_{12}(t)$ by assuming that ${\cal T}^{(n)}_{123}(t)$ is given at all time by\footnote{This expression holds at initial time for a statistical ensemble of independent particles at zero or finite temperature. In this case, we have:
\begin{eqnarray}
T_{123} &=& R_1R_2R_3 (1-P_{12}) (1- P_{13}-P_{23}) \nonumber \\
&=& D_{12} R_3 (1- P_{13}-P_{23}). \nonumber 
\end{eqnarray} 
}:
\begin{eqnarray}
{\cal T}^{(n)}_{123}(t) &=& {\cal D}^{(n)}_{12}(t) R^{(n)}_1(t) (1-P_{13} - P_{23}).  \label{eq:t123app}
\end{eqnarray}  
Using this expression in Eq. (\ref{eq:hybrif2temp}), we obtain that the equation of motion on ${\cal D}^{(n)}_{12}(t)$ can be recast as:
\begin{eqnarray}
i \hbar \frac{\partial {\cal D}^{(n)}_{12}}{\partial t} &=& \left[ h\left[ R_1^{(n)} \right] + h\left[ R_2^{(n)} \right], {\cal D}^{(n)}_{12} \right] \nonumber \\
&+& \frac{1}{2} \left(1 - R_1^{(n)} - R_2^{(n)}\right) \tilde{v}_{12} {\cal D}^{(n)}_{12} \nonumber \\
&-& \frac{1}{2} {\cal D}^{(n)}_{12} \tilde{v}_{12} \left(1 - R_1^{(n)} - R_2^{(n)} \right). \label{eq:hybrid2}
\end{eqnarray}
This equation, together with Eq. (\ref{eq:hybrid1}) will be the EOMs we will use in the following and that will replace the 
mean-field propagation in the phase-space method. 

In order to generalize the SMF approach, we still need to specify the statistical properties to be used for  $R_1^{(n)}(t_0)$ and ${\cal D}^{(n)}_{12}(t_0)$.  
One of our targeted goals is to fulfill the three requirements given by Eq. (\ref{eq:initrdt}). In particular the matching of the initial three-body density 
is not possible in the original phase-space approach when the Gaussian assumption is made on the initial ensemble.  A natural generalization 
would be to assume that 
\begin{eqnarray}
{\cal D}^{(n)}_{12}(t)  &=& D^{(n)}_{12}(t) + \Delta^{(n)}_{12}(t), \label{eq:alter}
\end{eqnarray}
where $D^{(n)}_{12}(t) $ can be for instance given by expression (\ref{eq:dsmf}) while $\Delta^{(n)}_{12}(t)$ has statistical properties chosen to insure 
at time $t_0$ that the second and third equations in (\ref{eq:initrdt}) are respected. One actually can also try to impose 
simultaneously that ${\rm Tr}_2 ({\cal D}^{(n)}_{12}(t) ) = (N-1)  R^{(n)}_{1}(t)$. This implies automatically ${\rm Tr}_2 (\Delta^{(n)}_{12}(t) ) =0$ 
at all time. We explored this strategy and tried to find a convenient statistical initial ensemble for $\Delta^{(n)}_{12}(t)$ with one or several of these constraints,   
but did not found any simple way.   
 
In the absence of a clear prescription, we finally simplified the problem and assumed that $R^{(n)}_1$ has the same initial statistical property as before given by Eqs. (\ref{eq:SMFfluc})
while the quantity ${\cal D}^{(n)}_{12}(t) $ is not fluctuating  initially  with:
\begin{eqnarray}
{\cal D}^{(n)}_{12}(t_0) &=& D_{12}(t_0) \label{eq:d12nf}.
\end{eqnarray} 
for all events. Then each initial condition is propagated using the equations (\ref{eq:hybrid1}) and (\ref{eq:hybrid2}). It should be noted in particular 
that, although ${\cal D}^{(n)}_{12}(t)$ is not fluctuating at initial time, it should be labelled by ${(n)}$ due to the initial fluctuations of ${R^{(n)}_1}$ that is used in 
Eq. (\ref{eq:hybrid2}).  In the absence of fluctuation on ${\cal D}_{12}$ at $t_0$ and with the condition (\ref{eq:d12nf}), it is immediate to verify that the two first constraints in (\ref{eq:initrdt}) are fulfilled while for the third one we have:
\begin{eqnarray}
\overline{{\cal T}^{(n)}_{123}}(t_0) &=& D_{12}(t_0) \overline{R^{(n)}_3(t_0)} (1-P_{13} - P_{23}), \nonumber \\
&=& D_{12}(t_0) R_3(t_0) (1-P_{13} - P_{23}). \nonumber 
\end{eqnarray} 
Therefore if $T_{123}(t_0)= D_{12}(t_0) R_3(t_0) (1-P_{13} - P_{23})$ in the initial conditions, the third constraint in (\ref{eq:initrdt}) is also fulfilled. This of course restrict 
the type of initial condition that could be considered. For instance, this will not allow to treat systems with initial residual non-zero three-body correlations. 
But systems that are initially described as a Slater determinant or a statistical ensemble of independent particles or eventually with only residual two-body correlations can be 
considered in the present approach.   
 
 \begin{figure}[!h]
    \includegraphics[width=1.0 \linewidth]{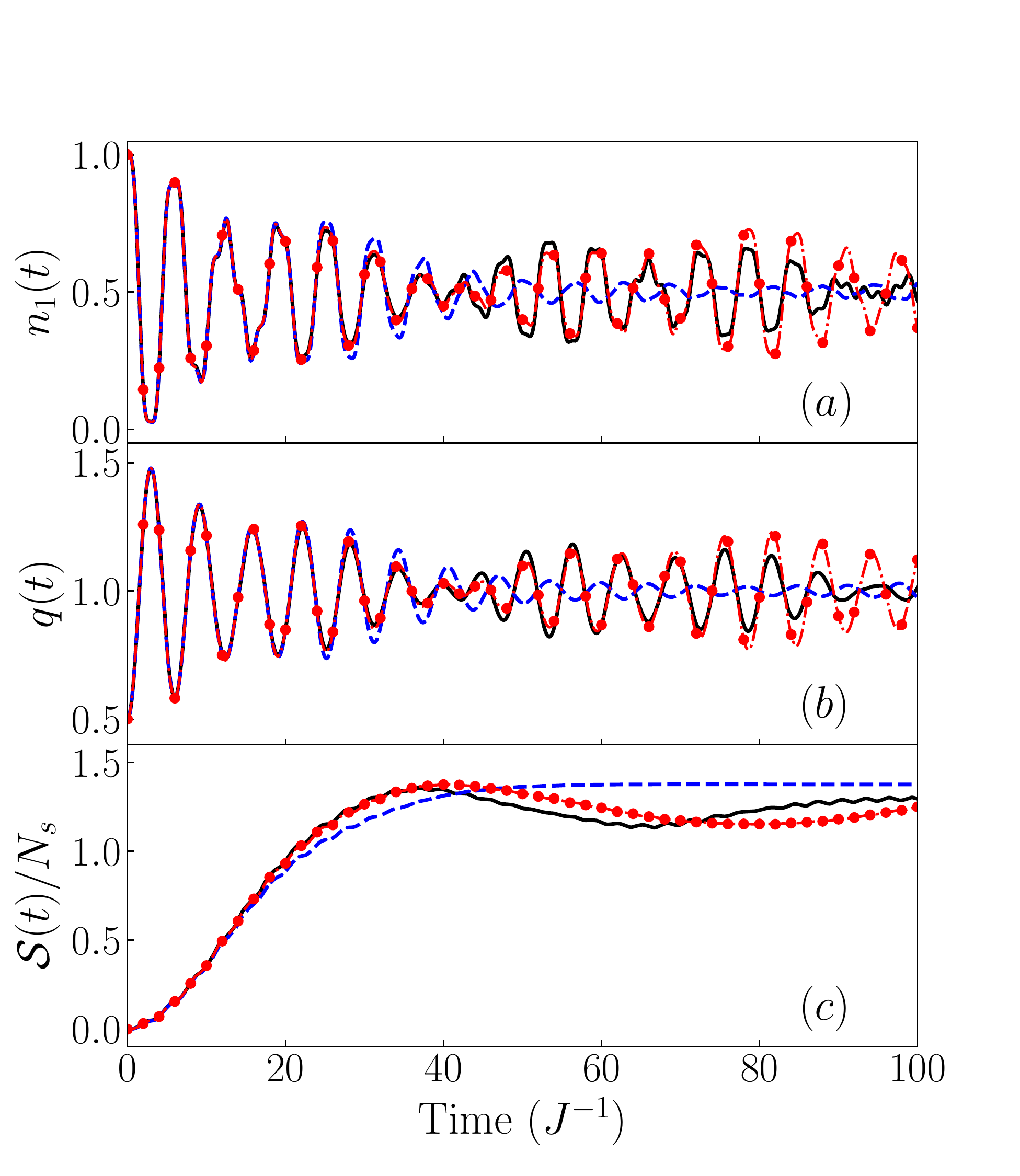}
    \caption{Time evolution of the (a) occupation probability of the leftmost site  (b) center of mass $q(t)$ of the interacting particles and (c) one-body entropy
    for  $U/J =0.1$ and $N= N_s = 4$ assuming that all particles are located on the left of the mesh initially. 
    In each panel, the exact solution is displayed by a black solid line, the results of the original SMF phase-space approach are shown by a blue dashed line. 
    The results of the HPS approach are shown with red filled circles.  
    In the SMF and HPS phase-space technique, results are obtained using 10000 trajectories. }
    \label{fig:N4U01}
\end{figure}

An important remark is that we keep the spirit of the SMF phase-space approach here. Indeed, all one-body quantities will be calculated using 
the equation (\ref{eq:o1}) and will be considered as classical objects. In particular fluctuations or equivalently correlations between   observables 
will still be performed using classical average over the sampled trajectories. Accordingly, as shown in the appendix \ref{sec:appfromsmfbbgky}, one can define 
a fluctuating two-body or three-body density ($D^{(n)}_{12}(t)$ or $T^{(n)}_{123}(t)$) along each path that are given by 
Eq. (\ref{eq:dsmf}) and (\ref{eq:tsmf}), and the only meaningful two-body density one could  extract  from the present formalism is the average of these quantities.  
In particular, ${\cal D}^{(n)}_{12}(t)$ obtained by solving the Eq. (\ref{eq:hybrid2}) or ${\cal T}^{(n)}_{123}(t)$ obtained by using  Eq. (\ref{eq:t123app}) should not be
confused in average with the two- and three-body densities obtained by the phase-space method. Note that, even if at initial time we have $\overline{{\cal D}^{(n)}_{12}(t_0)} =
\overline{{D}^{(n)}_{12}(t_0)} =  D_{12}(t_0)$, there is no reason that this equality is preserved for $t>t_0$.  We prefer to interpret these quantities as intermediate 
objects leading to a source term in Eq. (\ref{eq:hybrid1}) that has the effect to introduce effects beyond the mean-field.   

The present method, by using an initial statistical ensemble and where quantities are obtained by performing a classical statistical average clearly enters into the 
category of phase-space approaches. However, because we use intermediate quantities that do not fluctuate at initial time, we do not follow fully the strategy of 
the original SMF approach and for this reason we it will hereafter be called Hybrid Phase-Space (HPS) method.    

\section{Application of the HPS method}
\begin{figure}[!h]
    \includegraphics[width=1.0 \linewidth]{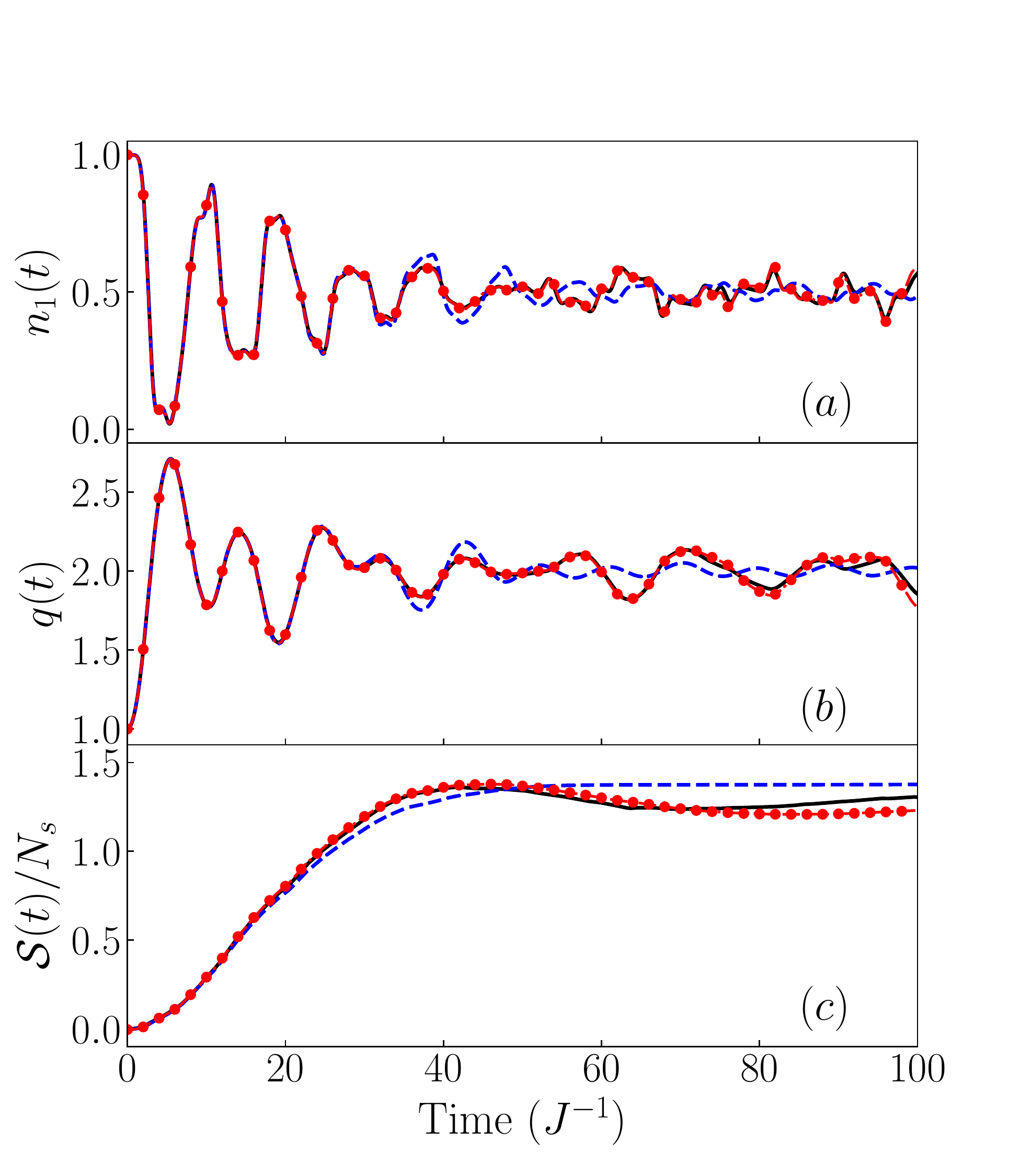}
    \caption{Same as figure \ref{fig:N4U01} for $N= N_s = 8$.  }
    \label{fig:N8U01}
\end{figure}    
In the present section, we apply the HPS method to the 1D Fermi-Hubbard model with different particle numbers and two-body interaction 
strengths. As we mentioned previously, this model is a perfect test-bench for improving the SMF phase-space method because, even in the weak coupling limit,
the SMF approach was not predictive for the long time evolution. For this model case, we give in appendix  \ref{sec:compdetHM} the explicit forms of the equations   
of motion that are used respectively for the SMF and for the HPS approaches as well as the properties of the initial fluctuations.   

\begin{figure}[!h]
    \includegraphics[width=1.0 \linewidth]{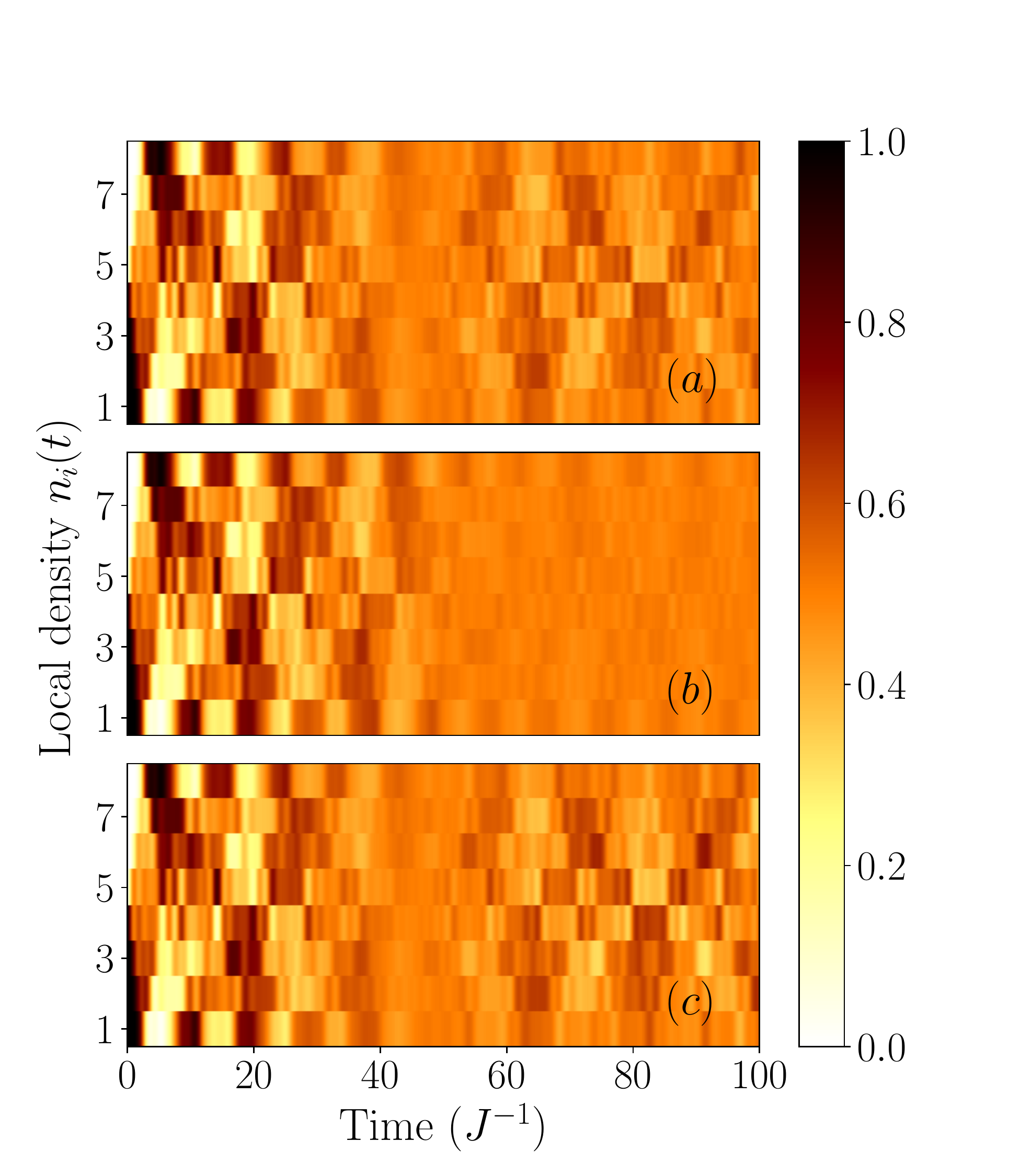}
    \caption{Time evolution of the  local density part of the one-body density $n_i(t)= R^{\sigma\sigma}_{ii}(t)$ for one of the spin orientation 
     as a function of time  obtained  for  $U/J =0.1$ and $N= N_s = 8$ assuming that all particles are located on one side of the mesh initially. The exact solution (a) 
    is compared to the SMF (b) and HPS (c) phase-space methods. }
    \label{fig:histo}
\end{figure}

We compare the exact evolution and approximate phase-space evolutions in Figs. \ref{fig:N4U01} and  \ref{fig:N8U01} obtained 
respectively for the case where $N= N_s = 4$ and $N= N_s = 8$ in the weak-coupling regime ($U/J = 0.1$) and when
all particles are located on one side of the mesh at initial time. Therefore, the initial condition in the mean-field consists in a Slater determinant
with initial spin symmetry.  In panel (a) of this figure, we display the occupation probability of the leftmost site. In the exact case, the occupation probability 
of the site $i$ verifies $n_{i\sigma}(t) = R^{\sigma \sigma}_{ii}$. Due to the initial condition, it verifies $n_{i\uparrow}(t) =  n_{i\downarrow}(t)$, allowing us to denote it simply by 
$n_i(t)$. In the phase-space approach, the occupation probability has the same spin symmetry and is defined through the average over events 
$n_i(t) = \overline{R^{\sigma \sigma(n)}_{ii}(t)}$.  
In panel (b) of these figures, we show a quantity $q(t)$ that could be interpreted as the equivalent to the center of mass of the particles. This quantity is defined 
as:
\begin{eqnarray}
q(t) &=& \frac{1}{2 N_s} \sum_{i,\sigma} \left( i- \frac{1}{2}\right) R^{\sigma\sigma}_{ii}(t). 
\end{eqnarray}
The factor $2$ comes from the fact that we sum over spins. 
 Finally in panel (c), we show the one-body entropy that is computed as:
\begin{eqnarray}
{\cal S}(t)  =  -k_B {\rm Tr}\left\{ R_1(t) \ln R_1(t) + (1-R_1(t)) \ln(1-R_1(t)) \right\}.  \nonumber
\end{eqnarray}
In practice, the entropy is computed by diagonalizing the average one-body density at time $t$. The entropy ${\cal S}(t)$ quantifies the departure from the pure Slater determinant case for which ${\cal S}(t)=0$.

\begin{figure}[!h]
    \includegraphics[width=1.0 \linewidth]{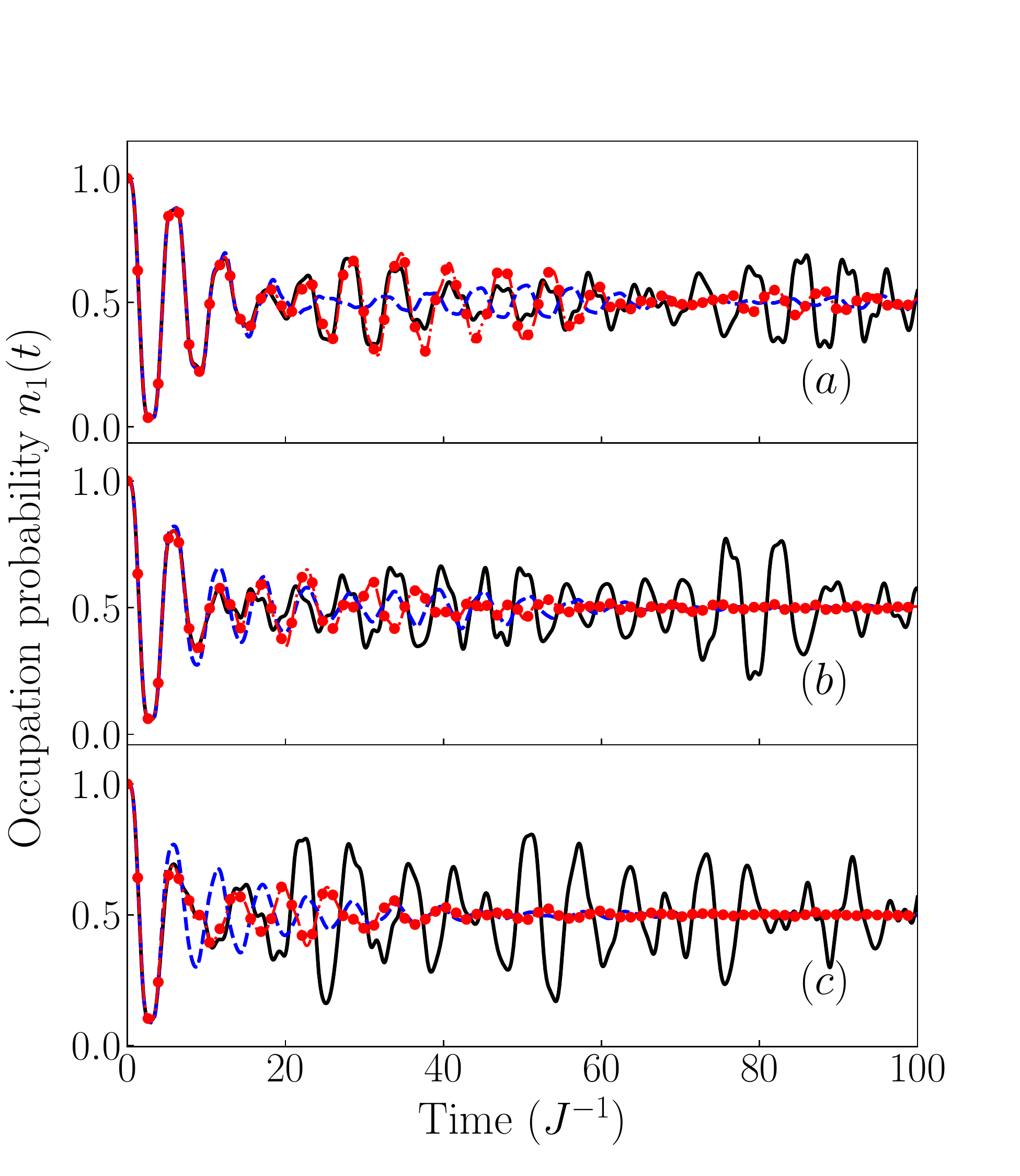}
    \caption{Time evolution of the occupation  probability of the leftmost site for $N= N_s = 4$ and different interaction strengths:  (a) $U/J =0.2$, (b) $U/J =0.4$ and (c)  $U/J =0.6$.
    In each case, all particles are initially located on one side of the mesh. 
    The exact solution is displayed using a black solid line, the result of the original SMF phase-space approach is shown by a blue dashed line and 
    the results of the HPS approach are shown with red filled circles.  }
    \label{fig:n1U020406N4}
\end{figure}

In Figures \ref{fig:N4U01} and  \ref{fig:N8U01}, we see that the new phase-space method proposed here is much better than the original SMF approach 
and not only reproduces the short time evolution but also the evolution over much longer time. In the case of weak coupling, we observe 
that the HPS evolution is almost on top of the exact evolution and only at very large time $U/J > 60$, very small deviations with the exact results are observed. 
In particular, the new phase-space approach does not suffer from the over-damping that is generally observed in SMF \cite{Lac12} and that is clearly seen 
in Fig.  \ref{fig:N4U01}.  By comparing the two figures, we also see that the agreement with the exact solution is improved when the number of particles increases. 

The fact that the long-time evolution is also reproduced by the new phase-space approach is quite surprising. 
Indeed, in the HPS approach as in the original SMF, the different trajectories are independent from each other. 
As shown in ref. \cite{Reg18,Reg19}, the long-time evolution of small systems can be treated in terms of a set of 
mean-field trajectories only if the quantum interferences between the trajectories are accounted for.     
\begin{figure}[!h]
    \includegraphics[width=1.0 \linewidth]{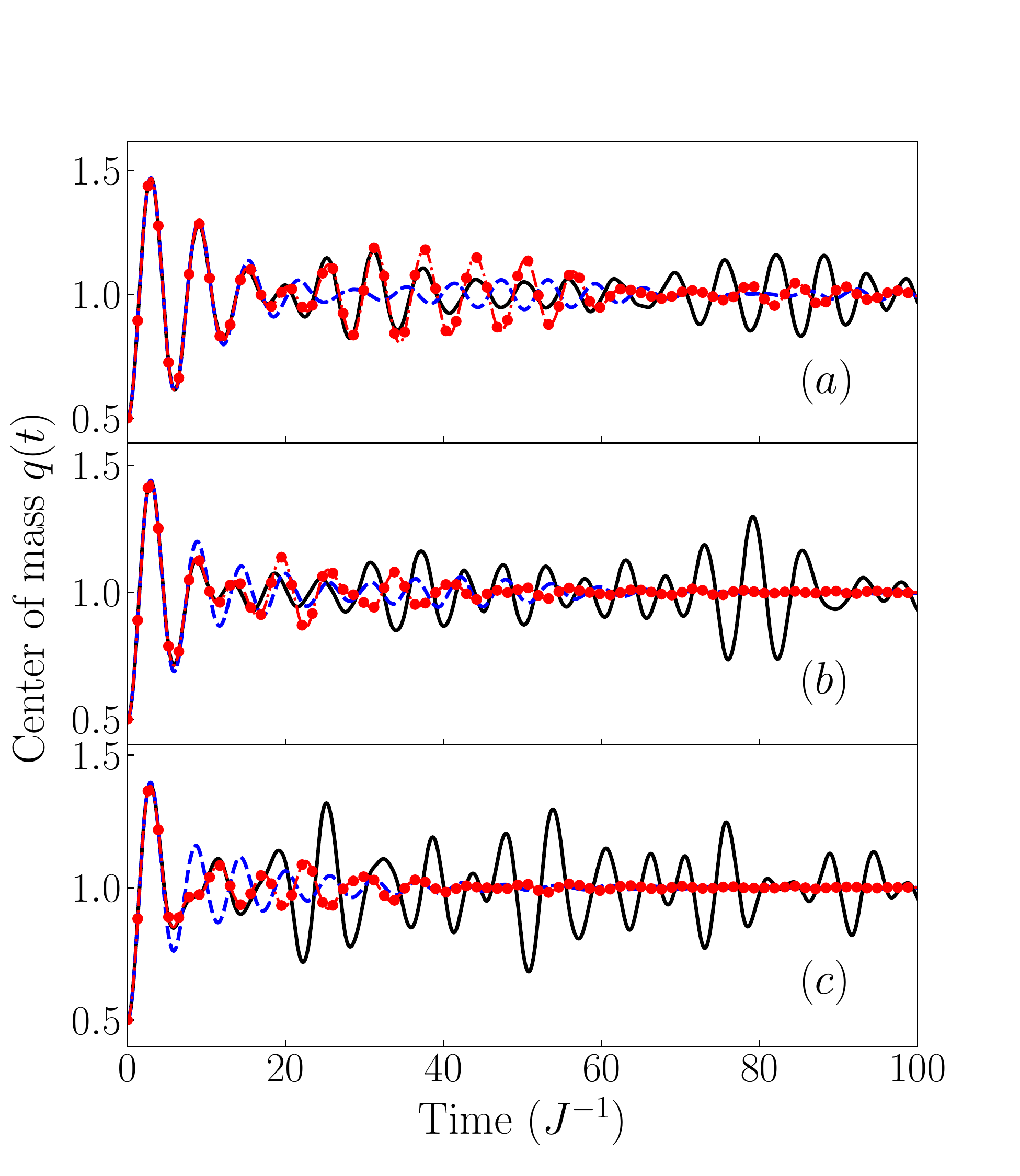}
    \caption{Same as figure \ref{fig:n1U020406N4} except that the center of mass motion $q(t)$ is shown instead of the leftmost site evolution.}
    \label{fig:qU020406N4}
\end{figure}
\begin{figure}[!h]
    \includegraphics[width=1.0 \linewidth]{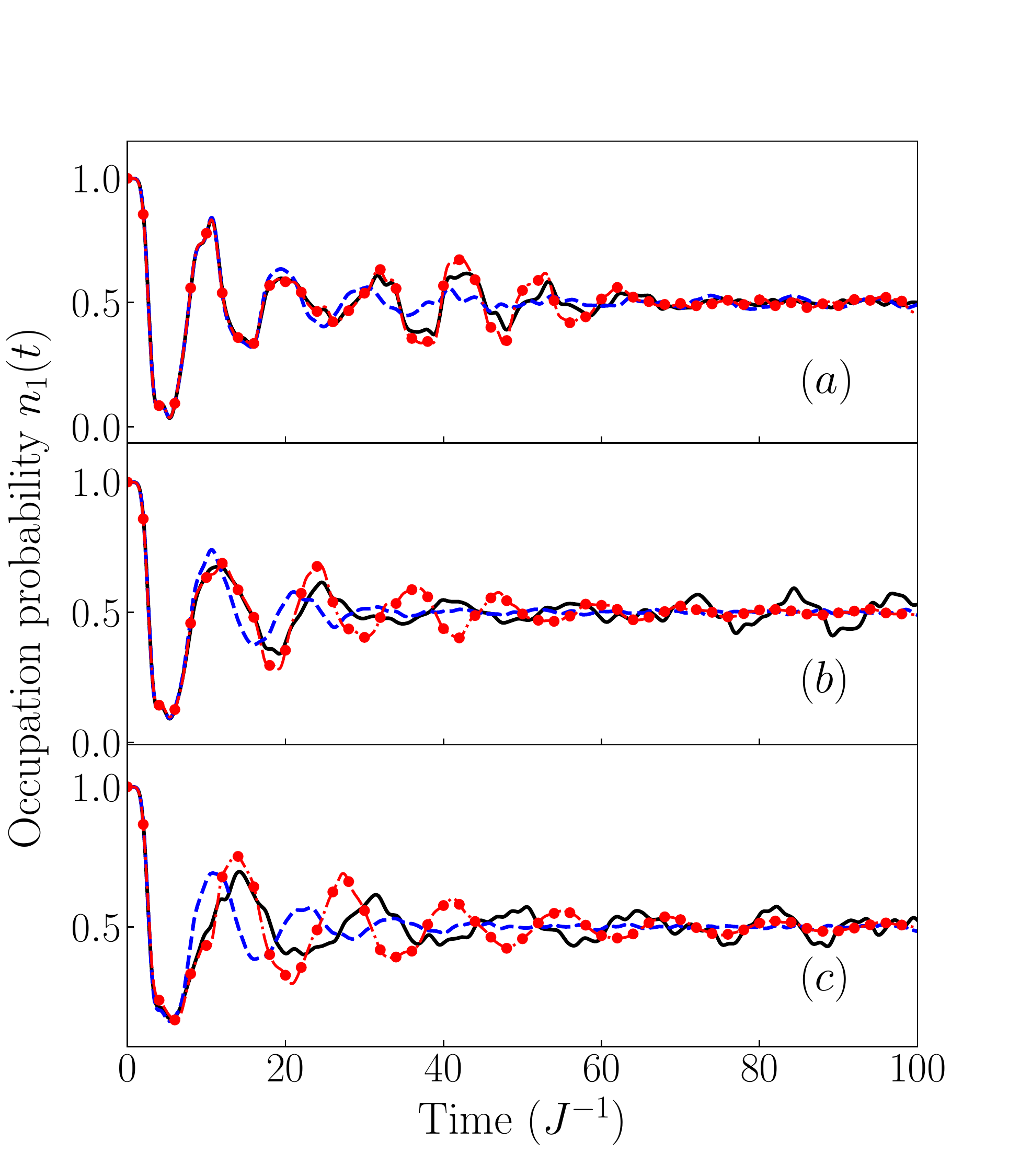}
    \caption{Same as Fig. \ref{fig:n1U020406N4} for $N= N_s = 8$. }
    \label{fig:n1U020406N8}
\end{figure}

\begin{figure}[!h]
    \includegraphics[width=1.0 \linewidth]{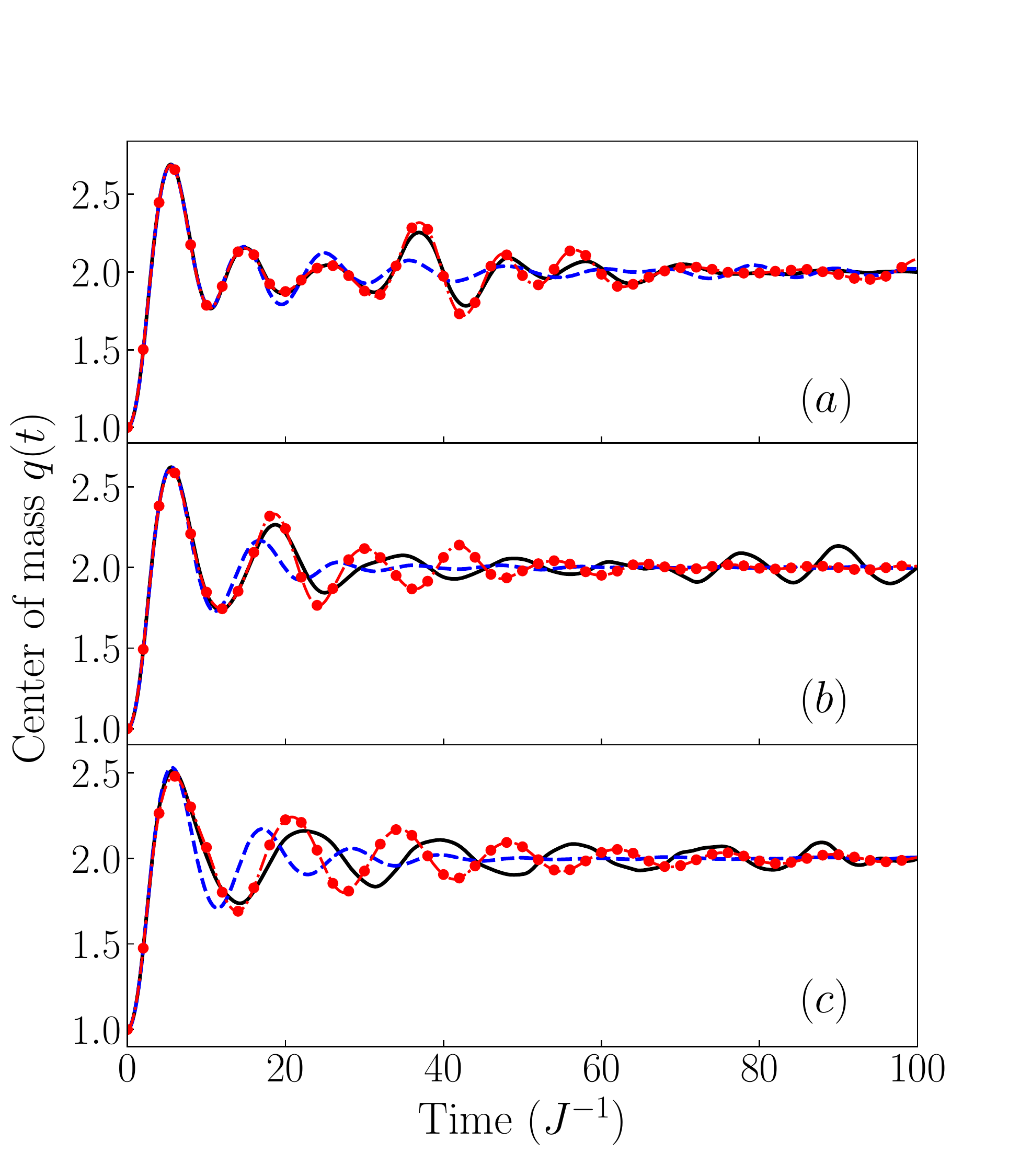}
    \caption{Same as Fig.  \ref{fig:n1U020406N4} except that the center of mass is now shown as a function of time for $N= N_s = 8$ 
    and varying interaction strength. }
    \label{fig:qU020406N8}
\end{figure}

\begin{figure}[!h]
    \includegraphics[width=1.0 \linewidth]{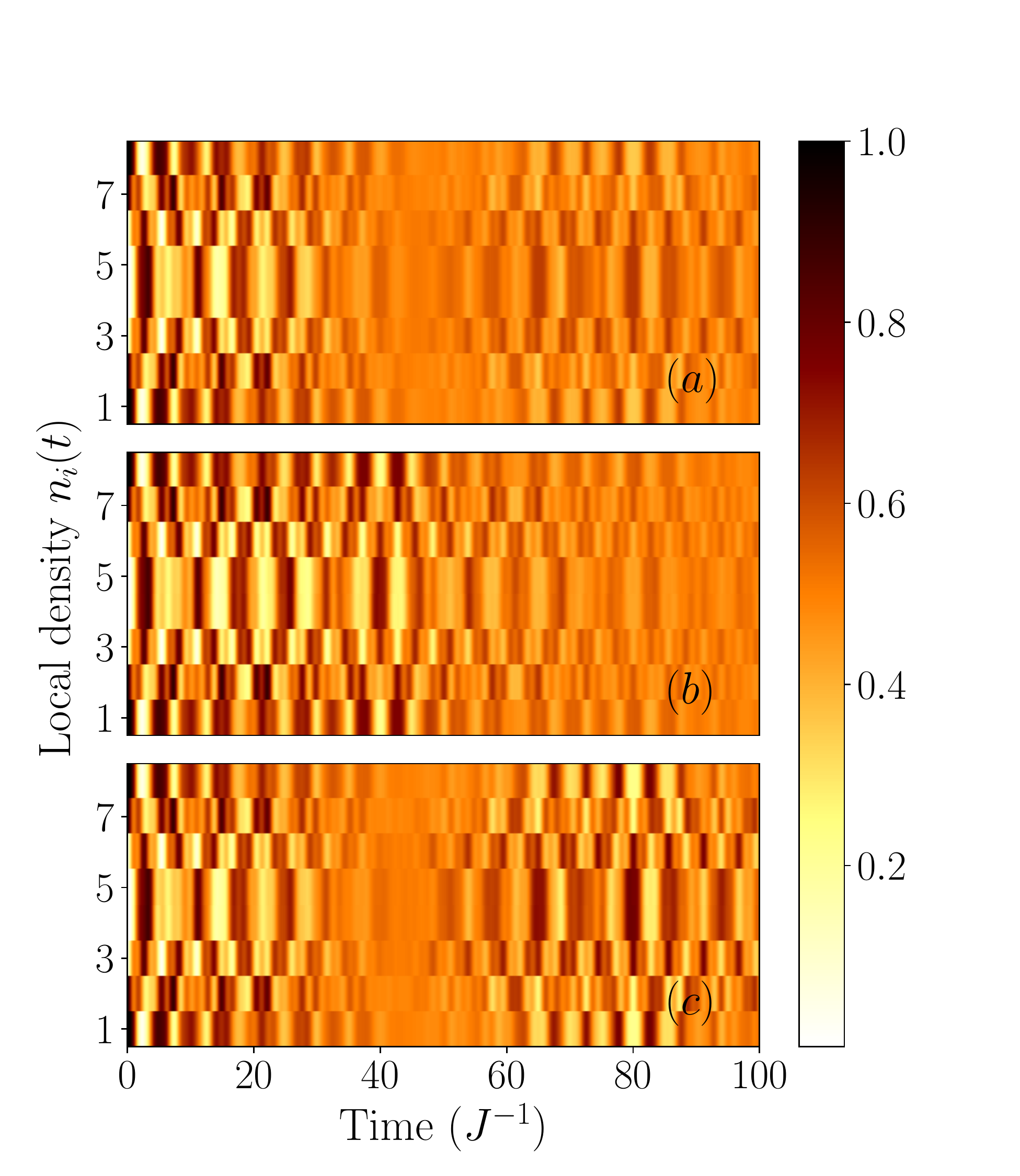}
    \caption{Same as figure \ref{fig:histo} except that the initial conditions are now two sets of particles located at each extremities of the lattice (see Eq. \eqref{eq:initialconditionscoll}). In this example, we assume that 4 particles are initially on the left and 4 on the right on a lattice of 8 sites.}
    \label{fig:histo_coll}
\end{figure}

Such interferences are indeed present in the Fermi-Hubbard model as illustrated in Fig. \ref{fig:histo}. In this figure, we show the evolution of the local density $n_i(t)$ 
as a function of time corresponding to the initial condition used in Fig.  \ref{fig:N8U01}. In this figure, the exact evolution seems to present interference patterns and revival 
of oscillations that are most probably due to the quantum wave that is bouncing back at the boundary. Such long time interference are not reproduced by SMF 
but are nicely reproduced in the HPS method.  This actually is a surprise in a method where trajectories are solved independently from each other. It should however 
be kept in mind that the HPS approximation goes beyond the independent particle motion by including part of the correlations that build up in time through the use of 
Eq. (\ref{eq:hybrid2}).   

In  Fig. \ref{fig:n1U020406N4} and Fig. \ref{fig:qU020406N4} for $N=4$ and Fig  \ref{fig:n1U020406N8} and Fig. \ref{fig:qU020406N8} for $N=8$, 
we show the evolution of the leftmost site occupation probability $n_1(t)$ and $q(t)$ respectively 
when the two-body coupling strength increases. In all cases, we observe that the HPS method reproduces much better the exact evolution than the SMF approach.  

However, when the two-body strength increases, we see after some time $\tau_{\rm HPS}$ some deviations with the exact evolution.   
The time-scale over which HPS is predictive decreases as $U/J$ increases as clearly illustrated in Figs. \ref{fig:n1U020406N4} and \ref{fig:qU020406N4}.  

A similar observation can be made for the SMF approach with a time-scale $\tau_{\rm SMF}$ over which the approach is reproducing the exact evolution.   
We clearly see in these figures that whatever the coupling $U/J$ is, we have always $\tau_{\rm SMF} < \tau_{\rm HPS}$.

Finally, as a further illustration  of the complex correlations that were missing in the SMF and that could be grasped by the HPS method, we also tried slightly different initial conditions.
We assumed for $N=8$ particles  that initially half the particles (here 4) are on the left site of the lattice while the other half is located on the right side.  (see Eq. \eqref{eq:initialconditionscoll}). The dynamics can be seen as a minimal version for two colliding Fermi systems. We show in Fig. \ref{fig:histo_coll} the local density evolution 
for the weak coupling regime with $U/J=0.1$. We compare in this figure the exact evolution (a) with the SMF (b) and HPS (c) results. 
The most striking feature is that while HPS catches on the exact dynamics up to intermediate time ($ 50 J^{-1}$) and then shows an underdamping of the oscillations in comparison with the exact dynamics, SMF deviates significantly from the exact case for $t \ge 20-25 J^{-1}$. We see with this figure the increase of predictive power of the HPS approach 
compared to the original phase-space method. 

Our conclusion is therefore that the novel phase-space method has globally 
a much better predictive power than the original phase-space approach based on the mean-field 
propagation. In particular, it seems extremely good in the weak coupling regime even for the long time evolution. 
The increase of predictive power, as discussed in section \ref{sec:hybrid}, can directly be traced back to the better account of the initial 
conditions with in particular the three-body density that is properly reproduced and a partial account for the two-body correlations 
in the evolution of each trajectory.  Note finally that we also applied the HPS to higher coupling strength ($U/J \geqslant 1$) but we observed 
that some trajectories are hard to converge unless very small numerical time-step are used. Therefore, in its present form, the HPS method 
is essentially restricted to weak- to medium-coupling regime.        
  
\section{Conclusion}

In this work, we explored the possibility to improve the predictive power of the SMF phase-space approach 
by relaxing the assumption that the equation of motion in this phase-space approach identifies with TDHF. Our strategy 
was to use the BBGKY hierarchy as a guidance and improve the evolution along each trajectory by including at least partially effects 
beyond the mean-field approximation. To do so, it was rather natural to us to assume that we consider not only a one-body density 
with initial fluctuations but also a two-body density that can fluctuate at initial time as proposed in Eq. (\ref{eq:alter}). 
Then, the two densities would follow a set of coupled equations that could be inspired from the TD2RDM approach. Unfortunately, 
the different attempt we made were unsuccessful and having both the one- and two-body densities that fluctuate lead to unstable 
trajectories preventing from performing the statistical average.

 We then propose here an alternative method where a set of one-body densities are still considered initially but where the TDHF approximation 
is corrected by an additional term that approximately describe the effect of correlations that built-up in time on the one-body evolution. This method mixes concepts taken from phase-space and BBGKY techniques and is called for this reason Hybrid Phase-Space approach. 
The applications of the novel approach to the one-dimensional Fermi-Hubbard model clearly demonstrates that the predictive power 
is improved compared to the original SMF technique. In particular, the new method is very effective in the weak-coupling regime and can even predict the long time 
evolution. This long-time evolution description was not possible with the original SMF technique. Overall, we see that the predictive power is increased for all coupling 
strength that are considered in this work. 

It should be noted that we observed in practice that the number of trajectories to be sampled in the HPS and SMF approach to obtain 
similar statistical errors are more or less the same. Still the numerical effort 
in the HPS approach is significantly increased due to the fact that the TDHF trajectory originally used in SMF is replaced by a TD2RDM like equation that is more numerically demanding .  Despite the extra numerical effort, the improved results obtained here are rather encouraging and the possibility to mix fluctuating with non-fluctuating initial conditions 
might open new perspectives.  

\newpage 
\appendix

\section{Equation of motion used for the Fermi-Hubbard Model}
\label{sec:compdetHM}

The EOMs in the Fermi-Hubbard model with sharp boundary conditions (see the Hamiltonian (\ref{eq:Hubbard_Hamiltonian})) can conveniently 
be written in the basis set of site orbitals with spin associated with the fermionic operators $(\hat c^\dagger_{i\sigma}, \hat c_{i\sigma})$. 
We denote by $N= N^\uparrow+ N^\downarrow$ the number of particles where $N^\uparrow$ (resp. $N^\downarrow$) is the number of particles with spin up (resp. down).
For $N_s$ sites, the size of the Hilbert N-body space  is given by $\dbinom{N_s}{N^\uparrow} \times \dbinom{N_s}{N^\downarrow}$. Some symmetries can eventually be used to reduce the numerical complexity of the problem: 
\begin{itemize}
\item The number of particles $N = \sum_i \left( n_{i \uparrow} + n_{i \downarrow} \right)$ is conserved, i.e.  $\left[ N,H \right]=0$,
\item The projection on the z-axis of the total spin $S_{z} = \frac{1}{2} \sum_{i} \left( n_{i \uparrow} - n_{i \downarrow} \right)$ is conserved: $\left[ S_{z}, H \right] = 0$.
\item As a consequence of the two symmetries above, the number of $+1/2$ particles and $-1/2$ particles are both conserved.
\end{itemize}

These symmetries imply that the Hamiltonian matrix will be block diagonal where a given block corresponds to a given value of $N$ and $S_z$ projection.
In particular, if the system has a given particle number and $S_z$ at initial time, its time-evolution only requires the corresponding part of the Hamiltonian in this sub-block, reducing 
significantly the numerical effort for the exact solution. 

The symmetries of the initial state that are preserved in time automatically implies some symmetries on the matrix elements of the one-, two-, $\cdots$ density matrices.     
Denoting the spin up (resp. spin down) with a $+$ (resp. $-$), and considering that the initial state corresponds to the $S_z=0$ (symmetry spin up/spin down) case, we have schematically:  
\begin{eqnarray}
R^{++} &=& R^{--} \nonumber
\\
R^{+-} &=& R^{-+} = 0 \nonumber
\\
D^{+-+-} &=& D^{-+-+} 
\\
D^{++++} &=& D^{----} = D^{+-+-} + D^{+--+} \nonumber
\\
D^{+--+} &=& D^{-++-} \nonumber
\end{eqnarray}
where $R$ and $D$ denote respectively the one and two-body density matrices (note that here the labels associated to site number are implicit). 
We can see that one only needs to propagate $R^{++}$ or $R^{--}$, and a careful analysis shows that $D^{+-+-}$ is the only component of the two-body density matrix that 
will affect the dynamics when propagating both the one-body and two-body degrees of freedom in the BBGKY hierarchy. Note that the quantity $\mathcal{D}^{(n)}_{12}$ introduced in this article follows the same symmetry properties as $D_{12}$.

We give below the different EOMs that are used in the present work (with the convention $\hbar =1$):

Omitting the spin indices on $R$ for clarity since no confusion can be made, and considering that the latin subscript $i,j,\dots$ denotes the $i^{th},j^{th},\dots,$ site starting from the left of the 1D lattice, one can write the EOMs for the TDHF, SMF and HPS theories: 
\begin{widetext}
\begin{itemize}
\item Mean-field EOM -- Assuming spin symmetry at initial time and using the notations $R_{ij} = R^{++}_{ij} =R^{--}_{ij}$, the TDHF evolution is given by:   
\begin{eqnarray}
i \dot{R}_{ij} &=& -J \left(R_{i+1j} (1-\delta_{iN_s}) + R_{i-1j} (1-\delta_{i1}) - R_{ij+1} (1-\delta_{jN_s}) - R_{ij-1} (1-\delta_{j1}) \right) + U R_{ij} \left(R_{ii} - R_{jj}\right).
\end{eqnarray}
Assuming that all particles are located on the left side of the lattice, the initial density is given by:
\begin{equation}
  R_{ij} (t_0)=\left\{
  \begin{array}{@{}ll@{}}
    1 & \text{if}\ i=j \ \text{and}\ i \leq N_s = N \\
    \\
    0 & \text{otherwise}
  \end{array}\right.
  \label{eq:initialconditions}
\end{equation} 
In another test, the initial conditions were modified to simulate the collision of two groups of particles of equal sizes initially 
disposed on each extremities of the mesh:
\begin{equation}
  R_{ij} (t_0)=\left\{
  \begin{array}{@{}ll@{}}
    1 & \text{if}\ i=j \ \text{and}\ i \not\in \left[N^\uparrow/2,N_s-N^\uparrow/2 \right] \\
    \\
    0 & \text{otherwise}
  \end{array}\right.
  \label{eq:initialconditionscoll}
\end{equation} 
\item SMF EOM -- In the original SMF phase-phase approach, the EOM remains the TDHF one except that the initial density is fluctuating at initial time. We then have:  
\begin{eqnarray}
i \dot{R}^{(n)}_{ij} &=& -J \left(R_{i+1j}^{(n)} (1-\delta_{iN_s}) + R_{i-1j}^{(n)} (1-\delta_{i1}) - R_{ij+1}^{(n)} (1-\delta_{jN_s}) - R_{ij-1}^{(n)} (1-\delta_{j1}) \right) + U R_{ij}^{(n)} \left(R_{ii}^{(n)} - R_{jj}^{(n)}\right),
\end{eqnarray}
where the initial at initial time:
\begin{eqnarray}
R_{ij}^{(n)} (t_0) &=& \overline{R_{ij}^{(n)} (t_0)} + \delta R_{ij}^{(n)} (t_0), 
\\
\overline{R_{ij}^{(n)} (t_0)} &=& R_{ij} (t_0) \nonumber.
\end{eqnarray}
The properties of $\delta R_{ij}^{(n)} (t_0)$ are specified in section \ref{sec:phasespace}. We would like to mention that we assume in the present SMF application as well as in the HPS presented below that spin up-spin down symmetry is respected along each path. Fluctuations that break the spin symmetry at initial time are allowed by the statistical properties of the one-body density $R_{ij}^{(n)}$ within SMF. For the SMF, this was tested 
and discussed in Ref. \cite{Lac14b}. The conclusion is that allowing the breaking of spin symmetry at initial time increases the numerical effort while not 
increasing/decreasing the predicting power. For this reason, we consider here the case where the spin symmetry is respected event-by-event.  

\item The HPS EOM --  In the HPS equation of motion, only $\mathcal{D}^{+-+-(n)}$ is coupled to $R^{(n)} = R^{++(n)} =R^{--(n)}$. For this reason, we use the compact notations $\mathcal{D}_{ijkl}^{(n)} = \mathcal{D}_{ijkl}^{+-+-(n)}$. The EOMs then read    
\begin{eqnarray}
i \dot{R}^{(n)}_{ij} &=& -J \left(R_{i+1j}^{(n)} (1-\delta_{iN_s}) + R_{i-1j}^{(n)} (1-\delta_{i1}) - R_{ij+1}^{(n)} (1-\delta_{jN_s}) - R_{ij-1}^{(n)} (1-\delta_{j1}) \right) + U \left( \mathcal{D}_{iiji}^{(n)} - \mathcal{D}_{ijjj}^{(n)} \right), \nonumber
\\
i \dot{\mathcal{D}}_{ijkl}^{(n)} &=& - J \left(\mathcal{D}_{i+1jkl}^{(n)} (1-\delta_{iN_s}) + \mathcal{D}_{i-1jkl}^{(n)} (1-\delta_{i1}) + \mathcal{D}_{ij+1kl}^{(n)} (1-\delta_{jN_s}) + \mathcal{D}_{ij-1kl}^{(n)} (1-\delta_{j1}) \right. \nonumber
\\
&& \left. - \mathcal{D}_{ijk+1l}^{(n)} (1-\delta_{kN_s}) - \mathcal{D}_{ijk-1l}^{(n)} (1-\delta_{k1}) - \mathcal{D}_{ijkl+1}^{(n)} (1-\delta_{l N_s}) - \mathcal{D}_{ijkl-1}^{(n)} (1-\delta_{l1})  \right) 
\\
&& + U \left(R_{ii}^{(n)} + R_{jj}^{(n)} - R_{kk}^{(n)} - R_{ll}^{(n)} \right) \mathcal{D}_{ijkl}^{(n)} \nonumber
\\
&& + U \left( \delta_{ij} \mathcal{D}^{(n)}_{iikl} - R_{ij} \mathcal{D}^{(n)}_{jjkl} - R_{ji} \mathcal{D}^{(n)}_{iikl}\right) -  U \left( \delta_{kl} \mathcal{D}^{(n)}_{ijkk} - R_{kl} 
\mathcal{D}^{(n)}_{ijkk} - R_{lk} \mathcal{D}^{(n)}_{ijll}\right). \nonumber
\end{eqnarray}
For an initial state that corresponds to a Slater determinant, we have the initial conditions: 
\begin{eqnarray}
R_{ij}^{(n)} (t_0) &=& \overline{R_{ij}^{(n)} (t_0)} + \delta R_{ij}^{(n)} (t_0),  \nonumber 
\\
\overline{R_{ij}^{(n)} (t_0)} &=& R_{ij} (t_0) \nonumber,
\\
\overline{\mathcal{D}_{ijkl}^{(n)} (t_0)} &=& R_{ik} (t_0) R_{jl} (t_0).\nonumber 
\end{eqnarray}
\end{itemize}

\section{General remark on SMF and some properties}
\label{sec:appfromsmfbbgky}

In Ref. \cite{Lac15}, it has been shown that the SMF approach can be linked to a hierarchy of equations 
of the moments of the one-body density that resembles  the BBGKY hierarchy. In the present section, we precise the link 
between the moments and the SMF approach. In SMF, one-body observables are treated as classical fluctuating objects that are 
given along each trajectory by:
\begin{eqnarray}
A^{(n)}(t) = \sum_{ij} A_{ij} R^{(n)}_{ji}(t) \label{eq:o1}
\end{eqnarray}  
where $ R^{(n)}_{ji}(t)$ are the densities with initial  fluctuations followed by TDHF evolution.

The SMF approach makes a mapping between quantum expectation values and classical statistical average. More precisely,  let us consider a set of 
one-body operators, denoted by $\hat A$, $\hat B$, $\hat C$, ... The following mapping is made:
\begin{eqnarray}
\begin{array}{lcl}
\displaystyle \langle \hat A \rangle      &  \longrightarrow &  \overline{A^{(n)}} = \sum_{ij } A_{ij}  \overline{R^{(n)}_{ji}},\\
\\
\displaystyle \langle \{ \hat A,  \hat  B \}_+ \rangle        &  \longrightarrow & \overline{A^{(n)} B^{(n)}} = \sum_{ij kl} A_{ij} B_{kl} \overline{R^{(n)}_{ji} R^{(n)}_{lk}}, \\
 \\
\displaystyle \langle \{  \hat  A,  \hat  B ,  \hat  C \}_+ \rangle        &  \longrightarrow & \overline{A^{(n)} B^{(n)}C^{(n)}} = \sum_{ij kl mn} A_{ij} B_{kl} C_{mn} \overline{R^{(n)}_{ji} R^{(n)}_{lk}R^{(n)}_{nm}},  \\
\\
&\cdots& \nonumber
\end{array}
\end{eqnarray}   
where we have used the notation:
\begin{eqnarray}
\langle \{  \hat  A,  \hat  B \}_+ \rangle &\equiv& \frac{1}{2} \langle  \hat A \hat B +  \hat B \hat A \rangle  \nonumber \\
\langle \{ \hat  A,  \hat B,  \hat  C\}_+ \rangle &\equiv&    \frac{1}{6} \langle  \hat A \hat B \hat C +  \hat  A \hat C \hat B  +  \hat B \hat A \hat C+ \hat  B \hat C \hat A +
 \hat  C \hat B \hat A+  \hat C \hat A \hat  B \rangle \nonumber \\
&\cdots& \nonumber
\end{eqnarray}
The above quantum average can be connected to the one-, two- and higher order many-body densities simply by setting $\hat A =   \hat{N}_{ji}$, $\hat B= \hat{N}_{lk}$, $\hat C =  \hat{N}_{nm}$ where we have introduced the notations $\hat N_{ij} = a^\dagger_j a_i$. A lengthy but straightforward calculation gives:
\begin{eqnarray}
R_{ij} &=& \langle N_{ij} \rangle, \label{eq:d1sym} \\
D_{ik,jl} &=& \langle \{ \hat N_{ij}, \hat N_{kl} \}_+ \rangle -  \frac{1}{2} \left( \delta_{il}  R_{kj} +  \delta_{kj} R_{il}  \right) , \nonumber \\
T_{jln;ikm} &=&   \langle \{  \hat{N}_{ji} ,\hat{N}_{lk}, \hat{N}_{nm} \}_+ \rangle  - \frac{1}{2} \left( \delta_{jk} D_{ln;im} + \delta_{lm} D_{nj;ki} + \delta_{jm} D_{nl;ik} + \delta_{li} D_{jn;km} + \delta_{ni} D_{jl;mk} + \delta_{nk} D_{lj,mi} \right) \label{eq:d2sym}
\\
&& - \frac{1}{6} \left( \delta_{jk} \delta_{lm} R_{ni} + \delta_{li} \delta_{jm} R_{nk} + \delta_{lm} \delta_{ni} R_{jk} + \delta_{nk} \delta_{li} R_{jm} + \delta_{ni} \delta_{jk} R_{lm} + \delta_{jm} \delta_{nk} R_{li} \right) \label{eq:d3sym} \nonumber, \\
&\cdots&  \nonumber 
\end{eqnarray}
Where $R_1$, $D_{12}$ and $T_{123}$ denote the one-, two-, three-body density matrix respectively. We see in particular that the  information content of the symmetric moments $\langle N_{ij} \rangle$,  $\langle \{ \hat N_{ij}, \hat N_{kl} \}_+ \rangle$, $\langle \{  \hat{N}_{ji} ,\hat{N}_{lk}, \hat{N}_{nm} \}_+ \rangle$ , ... is equivalent to the information content of the one-, two-, three-body, ... density matrix. 

These relationships on the quantum densities and quantum symmetric moments and the mapping between these moments and the density $R^{(n)}$ show that 
the equivalent of the two-, three- ... body density can also be constructed in the SMF theory. Based on the above relationships, we introduce the matrices 
$D^{(n)}_{12}$, $T^{(n)}_{123}$,... that are defined from the quantity $R^{(n)}$ used in SMF using: 
 \begin{eqnarray}
D^{(n)}_{ik,jl} &=& R^{(n)}_{ij} R^{(n)}_{kl} -  \frac{1}{2} \left( \delta_{il}  R^{(n)}_{kj} +  \delta_{kj} R^{(n)}_{il}  \right) , \label{eq:dsmf} \\
\nonumber \\
T_{jln;ikm}^{(n)} &=& + R_{ji}^{(n)} R_{lk}^{(n)} R_{nm}^{(n)} \nonumber \\
&-& \frac{1}{2} \left( \delta_{jk}  R^{(n)}_{il} R^{(n)}_{mn} 
+ \delta_{lm}  R^{(n)}_{kn} R^{(n)}_{ij} 
+ \delta_{jm} R^{(n)}_{in} R^{(n)}_{kl}
+ \delta_{li} R^{(n)}_{kj} R^{(n)}_{mn} 
+ \delta_{ni} R^{(n)}_{mj} R^{(n)}_{kl}
+ \delta_{nk}  R^{(n)}_{ml} R^{(n)}_{ij}
\right) \nonumber \\
&+& \frac{1}{3} \left( \delta_{jk} \delta_{lm} R_{ni}^{(n)} + \delta_{li} \delta_{jm} R_{nk}^{(n)} + \delta_{lm} \delta_{ni} R_{jk}^{(n)} + \delta_{nk} \delta_{li} R_{jm}^{(n)} + \delta_{ni} \delta_{jk} R_{lm}^{(n)} + \delta_{jm} \delta_{nk} R_{li}^{(n)} \right)\label{eq:tsmf}, \\
\nonumber \\
&\cdots& \nonumber  
\end{eqnarray}
\subsection*{Properties of the density matrices}

The density matrices $D^{(n)}$ and $T^{(n)}$ defined in Eq. (\ref{eq:dsmf}) and (\ref{eq:tsmf}) do automatically fulfill some important properties. 
For instance, after a rather lengthy but straightforward calculation, it is possible to show that we have \footnote{Note that we did not 
check for higher-order densities but we anticipate that similar relations holds}:
\begin{eqnarray}
{\rm Tr} R^{(n)}_1(t) &= & N, \nonumber \\
(N-1) {\rm Tr}_2 D^{(n)}_{12}(t) &=& R^{(n)}_1(t), \nonumber \\
(N-2) {\rm Tr}_3 T^{(n)}_{123}(t) &=& R^{(n)}_{12}(t), \nonumber \\
&\cdots& \nonumber
\end{eqnarray}
These are important properties that holds for the exact evolution and are automatically fulfilled on an event-by-event  basis and therefore also hold when averaging over events. 
Such requirement are known to be a critical issue when performing TD$n$RDM calculations \cite{Lac15}. In SMF, the  statistical properties of the initial conditions are 
constructed to insure that the first and second moments of the quantum fluctuations match the one obtained through the statistical average. This automatically implies that 
we have the properties:
\begin{eqnarray}
\overline{R^{(n)}_1 }(t=0) &=& R_{1}(t=0) , \nonumber \\
\overline{D^{(n)}_{12}(t=0) } &=& D_{12}(t=0) .
\end{eqnarray} 
However, the three-body average density does not a priori match the quantum three-body density, especially if a Gaussian approximation is made for the initial statistical ensemble
(see for instance the discussion in \cite{Ulg19}) 

\end{widetext}

\end{document}